\pdfoutput=1
\documentclass[10pt,a4paper,english,journal]{IEEEtran}
\usepackage[T1]{fontenc}
\usepackage[latin9]{inputenc}
\usepackage{units}
\usepackage{amsmath}
\usepackage{amssymb}
\usepackage{stackrel}
\usepackage{graphicx}

\makeatletter

\pdfpageheight\paperheight
\pdfpagewidth\paperwidth

\providecommand{\tabularnewline}{\\}

\usepackage{xcolor,soul}
\sethlcolor{lightgray}

\makeatother

\usepackage{babel}
\begin{document}

\title{Analysis of Outage Latency and Throughput Performance in Industrial
Factory 5G TDD Deployments}

\author{\IEEEauthorblockN{Ali A. Esswie$^{1,2}$,\textit{ }and Klaus I. Pedersen\textit{$^{1,2}$
}\\
$^{1}$Nokia Bell-Labs, Aalborg, Denmark\\
$^{2}$Department of Electronic Systems, Aalborg University, Denmark}}

\maketitle
$\pagenumbering{gobble}$
\begin{abstract}
The fifth generation (5G) new radio supports a diversity of network
deployments. The industrial factory (InF) wireless automation use
cases are emerging and drawing an increasing attention of the 5G new
radio standardization groups. Therefore, in this paper, we propose
a service-aware time division duplexing (TDD) frame selection framework
for multi-traffic deployments. We evaluate the performance of the
InF network deployments with the state-of-the-art 3GPP modeling assumptions.
In particular, we consider the dynamic TDD mode along with optimized
uplink power control settings. Multi-traffic coexistence scenarios
are also incorporated such that quality of service (QoS) aware dynamic
user scheduling and TDD link selection are introduced. Extensive system
level simulations are performed in order to evaluate the performance
of the proposed solutions, where the proposed QoS-aware scheme shows
68\% URLLC outage latency reduction compared to the QoS-unaware solutions.
Finally, the paper offers insightful conclusions and design recommendations
on the TDD radio frame selection, uplink power control settings and
the best QoS-coexistence practices, in order to achieve a decent URLLC
outage latency performance in the state-of-the-art InF deployments. 

\end{abstract}

\section{Introduction}

The 5G new radio (5G-NR) supports multiple service classes such as
the ultra-reliable and low latency communications (URLLC), and the
enhanced mobile broadband (eMBB) {[}1{]}. The URLLC services require
stringent radio and reliability targets, i.e., one way radio latency
of 1 ms with $99.999\%$ success probability, where the eMBB applications
demand extreme data rates {[}2, 3{]}. The indoor factory automation
(InF) {[}4, 5{]} use cases are emerging where the 5G-NR cellular communications
are envisioned to replace the Ethernet-based interconnections. The
early 5G commercial roll-outs are expected over the unpaired spectrum
due to the available large free bandwidth {[}6, 7{]}. Therefore, the
time division duplexing (TDD) is vital for the 5G success. For TDD
deployments, base-stations (BSs) are able to dynamically change their
respective radio frame configurations in order to meet the time-varying
traffic demands. 

Although dynamic TDD systems offer greater flexibility of the network
resources in line with the directional traffic demands, the stringent
URLLC latency and reliability targets are highly challenging in those
networks {[}8{]}. This is attributed to: (a) the non-concurrent availability
of the downlink (DL) and uplink (UL) transmission opportunities, and
(b) the additional cross link interference (CLI) of BSs and user-equipment's
(UEs) with concurrent opposite transmission links. 

The achievable URLLC outage performance has been widely investigated
for the indoor deployments {[}9-11{]}, where the indoor office deployments
are mainly considered. Although, to the best of our knowledge, there
is a lack of prior art of the URLLC performance analysis in the InF
dynamic TDD deployments and with the corresponding channel modeling
and design assumptions. Furthermore, in {[}12{]}, authors investigate
the achievable radio outage latency in the time-sensitive communications,
where a tighter synchronization and on-time delivery of packets are
considered. In TDD deployments, The structure of the DL and UL link
switching of the TDD radio frame and the BS-BS CLI have been proved
to have a dominating impact on the URLLC outage radio latency {[}8{]}.
Therefore, a diversity of inter-BS TDD radio frame coordination schemes
are introduced in the open literature. In {[}13-15{]}, coordinated
DL beam-forming and receiver design are proposed in order to isolate
the subspace of the BS-BS CLI in the spatial domain from the useful
signal subspace. Moreover, smarter dynamic UE scheduling and optimized
power control {[}16{]} are essential to control the network CLI. Those
schemes typically require an inter-BS coordination signaling overhead,
e.g., for exchanging the UE-specific allocation information. 

Opportunistic TDD frame coordination schemes are also developed in
order to partially or fully avoid the network CLI with simpler processing
requirements and less coordination overhead. In {[}17{]}, a set of
TDD system optimizations, such as hybrid frame design and slot-aware
dynamic UE scheduling, is combined in order to offer CLI-free channels
for the UEs of the worst channel conditions. Furthermore, a semi-static
TDD adaptation algorithm {[}18{]} is proposed to avoid the network
CLI while offering a semi-static dynamicity of the network TDD radio
frame to traffic demands. Finally, a reinforcement-learning (RL) based
TDD frame optimization scheme {[}19{]} has been proposed to autonomously
optimize the BS-specific TDD frame selection in a distributed manner,
where the achievable learning gain offers a considerable URLLC performance
improvement compared to reactive TDD adaptation schemes. Therein,
two learning instances have been defined. The first learning instance
estimates the best DL to UL symbol ratio to adopt during a radio frame
where the second learning network seeks the best corresponding symbol
placement across the frame such that the latency statistics are minimized.

In this paper, we propose a QoS-aware TDD system framework for emerging
InF TDD deployments. This includes service-aware dynamic UE scheduling,
TDD radio frame selection criterion. We comprehensively evaluate the
achievable URLLC outage latency performance within such deployments,
in combination with the eMBB services. First, we investigate the impact
of the UL power control setting and CLI on the URLLC outage performance.
Secondly, joint URLLC and eMBB QoS coexistence scenarios are considered.
QoS-aware TDD link selection and dynamic UE scheduling are incorporated
to balance among the feasibility of a decent URLLC outage latency
performance and the achievable eMBB capacity. Finally, we adopt an
RL based solution to dynamically optimize the selection of the BS-specific
TDD frame configuration for different load regions. The presented
performance evaluations are obtained through extensive system level
simulations where the latest 3GPP modeling guidelines are followed.
The paper offers insightful recommendations of the optimized TDD system
design aspects for the InF deployments to fulfill the URLLC stringent
targets. 

This paper is organized as follows. Section II introduces the system
modeling. Section III presents the considered QoS-aware dynamic user
scheduling and TDD link selection strategy. Section IV discusses the
simulation methodology and the major performance evaluation of the
proposed solution. Finally, conclusions are drawn in Section VI. 

\section{System Model}

We consider an InF TDD network with $C$ cells, each is equipped with
$N$ antennas. As depicted by Fig. 1, the network deployment follows
the 3GPP modeling guidelines for InF networks {[}4, 5{]}. There are
$K=K^{\textnormal{dl}}+K^{\textnormal{ul}}$ uniformly-distributed
UEs per cell, where $K^{\textnormal{dl}}$ and $K^{\textnormal{ul}}$
imply the number of the DL and UL UEs per cell. Each UE is equipped
with $M$ antennas, and is assumed to request both DL and UL transmissions,
respectively. The URLLC service is modeled with the FTP3 traffic model
{[}20{]}, where the DL and UL URLLC packets are of a finite size $\textnormal{\ensuremath{\mathit{f}^{dl}}}$
and $\textnormal{\ensuremath{\mathit{f}^{ul}}}$ bits, respectively.
URLLC packets arrive at the transmitter according to a Poisson Arrival
Process with mean packet arrival rates of $\textnormal{\ensuremath{\lambda}}^{\textnormal{dl}}$
and $\textnormal{\ensuremath{\lambda}}^{\textnormal{ul}},$ in the
DL and UL directions, respectively. Therefore, the offered URLLC load
per cell in the DL direction is calculated by: $\varOmega^{\textnormal{dl}}=$$K^{\textnormal{dl}}\times\textnormal{\ensuremath{\mathit{f}^{dl}}}\times\textnormal{\ensuremath{\lambda}}^{\textnormal{dl}}$,
and in the corresponding UL direction as: $\varOmega^{\textnormal{ul}}=K^{\textnormal{ul}}\times\textnormal{\ensuremath{\mathit{f}^{ul}}}\times\textnormal{\ensuremath{\lambda}}^{\textnormal{ul}}$.
The total offered URLLC load is expressed as: $\varOmega=\varOmega^{\textnormal{dl}}+\varOmega^{\textnormal{ul}}$.
In this paper, we also assume the eMBB-URLLC coexistence scenarios
solely in the DL direction, where $k_{\textnormal{eMBB}}^{\textnormal{dl}}\subset K^{\textnormal{dl}}.$
The eMBB traffic is modeled by a constant bit rate (CBR) per each
eMBB UE {[}21{]}, i.e., emulates a broadband video streaming service.
Specifically, it implies finite-size eMBB packets $\rho-\textnormal{bits}$
which arrive at the transmitter with a constant arrival rate in time.
For those scenarios, the total offered load in DL direction is calculated
as: $\varOmega^{\textnormal{dl}}=\left(\varOmega^{\textnormal{dl}}\right)^{\textnormal{eMBB}}+\left(\varOmega^{\textnormal{dl}}\right)^{\textnormal{urllc}}$,
where $\left(\varOmega^{\textnormal{dl}}\right)^{\textnormal{eMBB}}$
is the eMBB offered load. 

\begin{figure}
\begin{centering}
\includegraphics[scale=0.85]{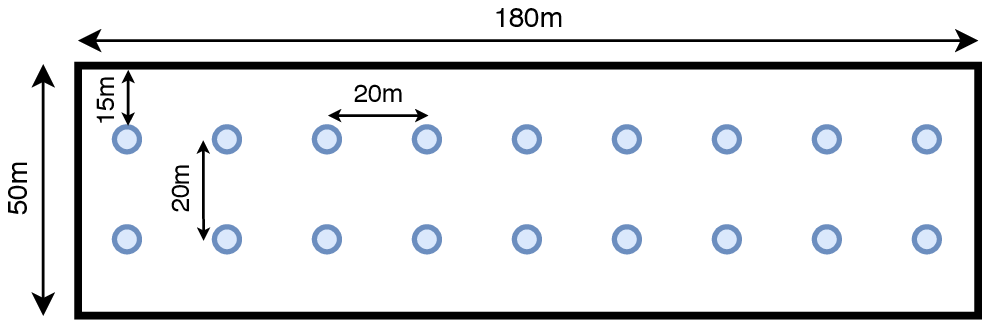}
\par\end{centering}
\centering{}{\small{}Fig. 1. System model: InF network deployment.}{\small \par}
\end{figure}

The UEs are dynamically multiplexed using the orthogonal frequency
division multiple access (OFDMA). In line with the 3GPP assumptions
for URLLC, we adopt a sub-carrier spacing (SCS) of 30 kHz with a physical
resource block (PRB) of twelve consecutive SCSs. We assume a short
transmission time interval (TTI) of 4 OFDM symbol duration for both
URLLC and eMBB transmissions. Before the start of each radio frame
(10 ms), each BS decides the structure of the selected slot formats
within the radio frame, where there is a guard symbol between each
DL and UL TTI transition to compensate for the channel propagation
delay. 

In line with the system modeling assumptions in {[}19{]}, we explicitly
consider the major functionalists of the 5G NR PHY and MAC layers.
In the DL direction, arriving packets are first processed, and are
buffered towards the first available DL TTI of the current TDD radio
frame. The DL UEs are dynamically scheduled using the adopted MAC
scheduler. The DL packets are subject to a further processing delay
at the UE-side. In case the DL packets are not successfully decoded,
UEs trigger a HARQ negative ACK (NACK) during the first available
UL TTI. Subsequently, the serving BSs re-transmit the failed DL packets
during the next available DL transmission opportunity. We adopt dynamic
link adaptation in the DL direction, based on periodic reporting of
the channel quality indications (CQIs) to select the best corresponding
modulation and coding scheme (MCS) to achieve the target block error
rate (BLER). 

In the UL direction, in line with {[}22{]}, we consider  the configured
grant (CG) transmission with a fixed MCS per UE. With CG, the arriving
UL packets at the UE-side side are immediately prepared for UL transmissions
during the first available UL TTI. This removes the delay for transmitting
the scheduling request until receiving the corresponding scheduling
grant. All CG-based UL transmissions include a robust preamble such
that the BS is able to distinguish from which UEs the UL transmission
is initiated. The CG UL configurations are set such that all active
UL UEs transmit over a randomly selected sub-band, of one quarter
of the carrier bandwidth, with a predefined MCS level of QPSK rate
$\nicefrac{1}{2}$. This setting allows for transmitting one full
URLLC packet in a single shot without segmentation.

The UL transmission power is configured as

\begin{equation}
\Sigma\left[dBm\right]=\min\left\{ \Sigma_{\textnormal{max}},\,\,P0+10\log_{10}\left(\text{\ensuremath{\wp}}\right)+\alpha\eth\right\} ,
\end{equation}
where $\Sigma_{\textnormal{max}}$ is the maximum UE transmit power,
$P0$ is the power spectral density, $\text{\ensuremath{\wp}}$ is
the number of granted UL PRBs, $\alpha$ and $\eth$ are the path-loss
compensation factor and path-loss. As CG transmissions from different
UEs can be transmitted over overlapping resources, those are subject
to intra-cell interference. In case the BS fails to decode UL transmissions
from different UL UEs, the BS triggers a re-transmission request during
the first available DL TTI with a dedicated scheduling grant for the
UE. Correspondingly, the UL UE initiates a packet re-transmission
using the same MCS and bandwidth configuration as the first UL transmission,
with a 3 dB transmission power boost to improve the decoding probability
of the HARQ re-transmission {[}22{]}. 

\section{Key Factors Impacting The URLLC Performance in InF Deployments }

In the following subsections, we show the critical 5G NR system design
aspects impacting the achievable URLLC outage performance within the
emerging InF TDD deployments. Those span the optimization of the UL
power control settings, dynamic UE scheduling, and the TDD link selection
framework, respectively. 

\subsection{QoS-aware TDD Radio Frame Selection }

In dynamic TDD networks, BSs independently select the radio frame
structures, in terms of the number of the DL and UL transmission opportunities
across the frame duration, that best meet their respective traffic
needs. Therefore, BSs continuously monitor their offered traffic demands
in the DL and UL directions. We formulate the relative buffered traffic
ratio $\mu_{[t,c]}\left(\varsigma\right)$ at the $\varsigma^{th}$
slot of the radio frame, $\varsigma=1,2,\ldots,\xi$, and $\xi$ is
the number of slots per the radio frame as 

{\small{}
\begin{equation}
\mu_{[t,c]}\left(\varsigma\right)=\frac{Z_{[t,c]}^{\textnormal{dl}}\left(\varsigma\right)}{Z_{[t,c]}^{\textnormal{dl}}\left(\varsigma\right)+\left(\nicefrac{1}{\iota}\right)\,Z_{[t,c]}^{\textnormal{ul}}\left(\varsigma\right)},
\end{equation}
}where $Z_{[t,c]}^{\textnormal{dl}}\left(\varsigma\right)$ and $\left(\nicefrac{1}{\iota}\right)Z_{[t,c]}^{\textnormal{ul}}\left(\varsigma\right)$
denote the total DL and UL buffered traffic size of the $\varsigma^{th}$
slot during the current frame, and $\iota$ implies the first-transmission
average UL BLER at the BS side. The latter is linearly averaged across
all UL transmissions and updated using a sliding window per UE. The
intuition of such formulation is derived by the fact that the BS has
different knowledge of the $Z_{[t,c]}^{\textnormal{dl}}\left(\varsigma\right)$
and $Z_{[t,c]}^{\textnormal{ul}}\left(\varsigma\right)$ information.
In particular, the knowledge of the $Z_{[t,c]}^{\textnormal{dl}}\left(\varsigma\right)$
is available at the BS. However, in the UL direction, the buffered
first UL transmission size per UE $Z_{[t,c]}^{\textnormal{ul}}\left(\varsigma\right)$
is not immediately accessible at the BS until it is received at the
BS side. Therefore, the term $\left(\nicefrac{1}{\iota}\right)Z_{[t,c]}^{\textnormal{ul}}\left(\varsigma\right)$
is adopted to reflect the actual offered UL traffic size, i.e., equivalent
to $Z_{[t,c]}^{\textnormal{dl}}\left(\varsigma\right)$ in the DL
direction. 

For multi-traffic deployments with joint URLLC-eMBB, the terms $Z_{[t,c]}^{\textnormal{dl}}\left(\varsigma\right)$
and $\left(\nicefrac{1}{\iota}\right)Z_{[t,c]}^{\textnormal{ul}}\left(\varsigma\right)$
represent the aggregate URLLC-eMBB buffered traffic sizes in the DL
and UL directions, respectively. As the eMBB traffic demand is typically
much larger than of the corresponding URLLC, the buffer ratio in (2)
and the selection of the TDD frame are both dominated by the eMBB
traffic statistics instead. This could be problematic to achieve a
decent URLLC outage latency due to the additional URLLC packet buffering,
i.e., due to the selection of a radio frame configuration that does
mainly satisfy with the buffered URLLC traffic. Therefore, we adopt
a QoS-aware TDD link selection criterion such as the buffered traffic
statistic in (2) is biased towards the URLLC QoS as follows

\begin{equation}
\begin{array}{c}
Z_{[t,c]}^{\textnormal{dl/ul}}\left(\varsigma\right)\rightarrow\left(Z_{[t,c]}^{\textnormal{dl/ul}}\left(\varsigma\right)\right)^{\textnormal{urllc}},\textnormal{URLLC-only}\\
Z_{[t,c]}^{\textnormal{dl/ul}}\left(\varsigma\right)\rightarrow\left(Z_{[t,c]}^{\textnormal{dl/ul}}\left(\varsigma\right)\right)^{\textnormal{eMBB}},\textnormal{eMBB-only}
\end{array},
\end{equation}
where $\left(Z_{[t,c]}^{\textnormal{dl}}\left(\varsigma\right)\right)^{\textnormal{urllc}}$,
$\left(Z_{[t,c]}^{\textnormal{dl}}\left(\varsigma\right)\right)^{\textnormal{eMBB}}$,
$\left(\nicefrac{1}{\iota}Z_{[t,c]}^{\textnormal{ul}}\left(\varsigma\right)\right)^{\textnormal{urllc}}$,
and $\left(\nicefrac{1}{\iota}Z_{[t,c]}^{\textnormal{ul}}\left(\varsigma\right)\right)^{\textnormal{eMBB}}$
are the aggregate DL and UL buffered traffic sizes for the URLLC and
eMBB UEs, respectively. The instantaneous buffered traffic ratios
$\mu_{[t,c]}\left(\varsigma\right)$ are averaged over the duration
of the TDD radio frame given as

{\small{}
\begin{equation}
\overline{\mu}_{[t,c]}=\frac{1}{\xi}\,\stackrel[\varsigma=1]{\xi}{\sum}\mu_{[t,c]}\left(\varsigma\right),
\end{equation}
}where $\overline{\mu}_{[t,c]}$ is the average traffic ratio of the
current radio frame. The traffic ratio $\overline{\mu}_{[t,c]}\rightarrow\left[0,1\right]$
implies the combined buffering performance of the DL and UL traffic
size. For example, $\overline{\mu}_{[t,c]}=0.1$ implies that the
buffered UL traffic is $9\textnormal{x}$ times the corresponding
DL traffic. Therefore, the corresponding BS shall select a TDD radio
frame with 90\% of time allocation to the UL transmission opportunities,
assuming a similar UL and DL spectral efficiency. The DL and UL symbols
of the selected radio frames are evenly distributed in terms of 4
OFDM symbol blocks, following the adopted DL-to-UL symbol ratio. 

\subsection{QoS-aware Dynamic UE Scheduling }

To highlight the impact of the UE scheduler, we adopt two frameworks
of the multi-QoS dynamic UE schedulers. First, we consider the well-known
weighted proportional fair (PF) criterion {[}23{]} to dynamically
schedule different URLLC and eMBB UEs in the time and frequency domains.
UEs are sorted in the time domain such as the URLLC UEs are always
given a higher priority than the eMBB UEs, i.e., URLLC UEs are given
a higher weight in the PF criterion. Therefore, the higher PF weight
of the URLLC UEs aims to always schedule the active URLLC UEs before
the respective eMBB UEs in the time domain. Thereafter, active URLLC
and eMBB UEs are both scheduled based on the PF criterion in the frequency
domain. That is, according to their achievable instantaneous throughput
relative to the total received capacity. This way, the scheduling
fairness is always guaranteed in the frequency domain among each set
of the URLLC and eMBB UEs, respectively. The main drawback of such
scheduling framework is that the URLLC latency statistics are not
considered in the scheduling criterion, and therefore, it could lead
to a degraded URLLC outage latency performance. 

Secondly, we adopt the scheduling framework introduced in {[}24{]}.
Instead of the throughput-based PF scheduling criterion, the head
of line delay (HoLD) is the basic scheduling criterion. The HoLD per
packet per UE is defined as the time from the DL  packet arrives at
the transmitter end until it is successfully decoded at the intended
receiver end. The scheduler always prioritizes an immediate scheduling
for the URLLC UEs with the largest HoLD statistics while requiring
the least packet segmentation. The intuition is that the scheduler
seeks to minimize the probability of the URLLC packet segmentation
probability, therefore, reducing the URLLC outage latency. In case
packet segmentation is not avoidable due to the resource shortage,
the scheduler seeks to segment a single URLLC packet that leads to
the minimum control overhead per TTI, hence, leaving more resource
for data transmissions. In joint URLLC-eMBB deployments, such scheduler
is proved to offer considerable eMBB capacity, due to the faster transmissions
of the concurrent URLLC packets, therefore, leaving more resources
for the corresponding eMBB traffic. 

\section{Performance Evaluation}

\subsection{Simulation Methodology }

We adopt a highly-detailed system level simulations to evaluate the
performance of the proposed solutions. The main set of the simulation
parameters is listed in Table I. We adopt the dense clutter - high
BS propagation model of the InF deployments {[}4, 5{]}, where the
BSs are elevated as compared to active UEs. The simulator used for
the system level evaluations has a timing resolution of a single OFDM
symbol and includes the main functionalities of the 5G NR protocol
stack. The simulator is validated via calibration exercises, where
baseline statistics for predefined simulation scenarios are reported
and compared between the various 3GPP partners {[}25{]}. For each
radio frame of 10 ms, the BSs select the radio frame configurations
which best suit their current DL and UL traffic demand. During the
DL TTIs, UEs are dynamically scheduled using either the PF or min-HoLD
{[}24{]} criterion. During the UL TTIs, UEs transmit their UL packets
using the CG UL following the settings presented in Section II. For
DL/UL packets, the signal to interference noise ratios (SINR) of the
granted sub-carriers are calculated using the linear minimum mean
square error interference rejection and combining receiver (L-MMSE-IRC).
Those are combined using the mean mutual information per coded bit
(MMIB) mapping {[}26{]} in order to estimate the effective SINR point.
Based on the effective SINR, the corresponding error probability is
calculated using look-up tables, obtained from extensive link level
simulations, considering the received effective SINR and the adopted
MCS. 

\begin{table}
\caption{{\small{}Simulation parameters.}}
\centering{}%
\begin{tabular}{c|c}
\hline 
Parameter & Value\tabularnewline
\hline 
Environment & 3GPP-InF, one cluster, 18 cells\tabularnewline
\hline 
UL/DL channel bandwidth & 20 MHz, SCS = 30 KHz, TDD\tabularnewline
\hline 
Channel model & InF-DH (dense clutter and high BS) {[}5{]}\tabularnewline
\hline 
BS and UE transmit power & BS: 30 dBm, UE: 23dBm\tabularnewline
\hline 
Carrier frequency & 3.5 GHz\tabularnewline
\hline 
BS and UE heights & BS: 10m, UE: 1.5m\tabularnewline
\hline 
Antenna setup & $N=4$ , $M=4$ \tabularnewline
\hline 
Average UEs per cell & $K^{\textnormal{dl}}=K^{\textnormal{ul}}=$ 8-16\tabularnewline
\hline 
TTI configuration & 4-OFDM symbols\tabularnewline
\hline 
URLLC Traffic model & $\begin{array}{c}
\textnormal{FTP3, \textnormal{\ensuremath{\mathit{f}^{dl}}} = \textnormal{\ensuremath{\mathit{f}^{ul}}} = 256 bits}\\
\textnormal{\ensuremath{\textnormal{\ensuremath{\lambda}}^{\textnormal{dl}}} = 50 pkts/sec}\\
\textnormal{\ensuremath{\textnormal{\ensuremath{\lambda}}^{\textnormal{ul}}} = 50 pkts/sec}
\end{array}$\tabularnewline
\hline 
eMBB Traffic model & CBR, $\rho=$ 16k bits, rate/UE = 0.5 Mbps\tabularnewline
\hline 
DL scheduling & PF, min-HoLD {[}24{]}\tabularnewline
\hline 
UL scheduling & CG, QPSK1/2, $P0$= -61 dBm, $\alpha$= 1\tabularnewline
\hline 
Processing time & $\begin{array}{c}
\textnormal{PDSCH prep. delay: 2.5-OFDM symbols}\\
\mathit{\textnormal{PUSCH prep. delay: 5.5-OFDM symbols}}\\
\textnormal{PDSCH decoding : 4.5-OFDM symbols}\\
\textnormal{PUSCH decoding: 5.5-OFDM symbols}
\end{array}$\tabularnewline
\hline 
DL/UL receiver & 	L-MMSE-IRC\tabularnewline
\hline 
TDD frame  & 10 ms\tabularnewline
\hline 
\end{tabular}
\end{table}

\subsection{Performance Results}

The UL power control settings have a vital impact on the overall URLLC
performance. Fig. 2 presents the complementary cumulative distribution
function (CCDF) of the achievable URLLC latency with different UL
power control settings, i.e., for several $P0$ configurations. As
can be clearly observed, with $P0=-90$ to $-60$ dBm, a decent URLLC
outage latency performance is obtained. Herein, the majority of the
UL UEs transmit their UL packets with a lower transmission power.
Therefore, the inter-cell interference is controlled while the UL
packet queuing delay dominates the achievable URLLC outage latency.
With very high $P0=-40$ to $-30$ dBm, the majority of the UL UEs
transmit their payload with the maximum permissible transmission power,
resulting in a significant increase of the inter-cell interference.
Hence, the interference starts to dominate the URLLC outage latency
where the packets require multiple HARQ re-transmission combing attempts
before a successful decode, leading to a highly degraded URLLC outage
latency performance. Based on the obtained URLLC performance in Fig.
2, we adopt $P0=-61$ dBm for the rest of the results in order to
achieve the best possible URLLC outage latency. 

\begin{figure}
\begin{centering}
\includegraphics[scale=0.6]{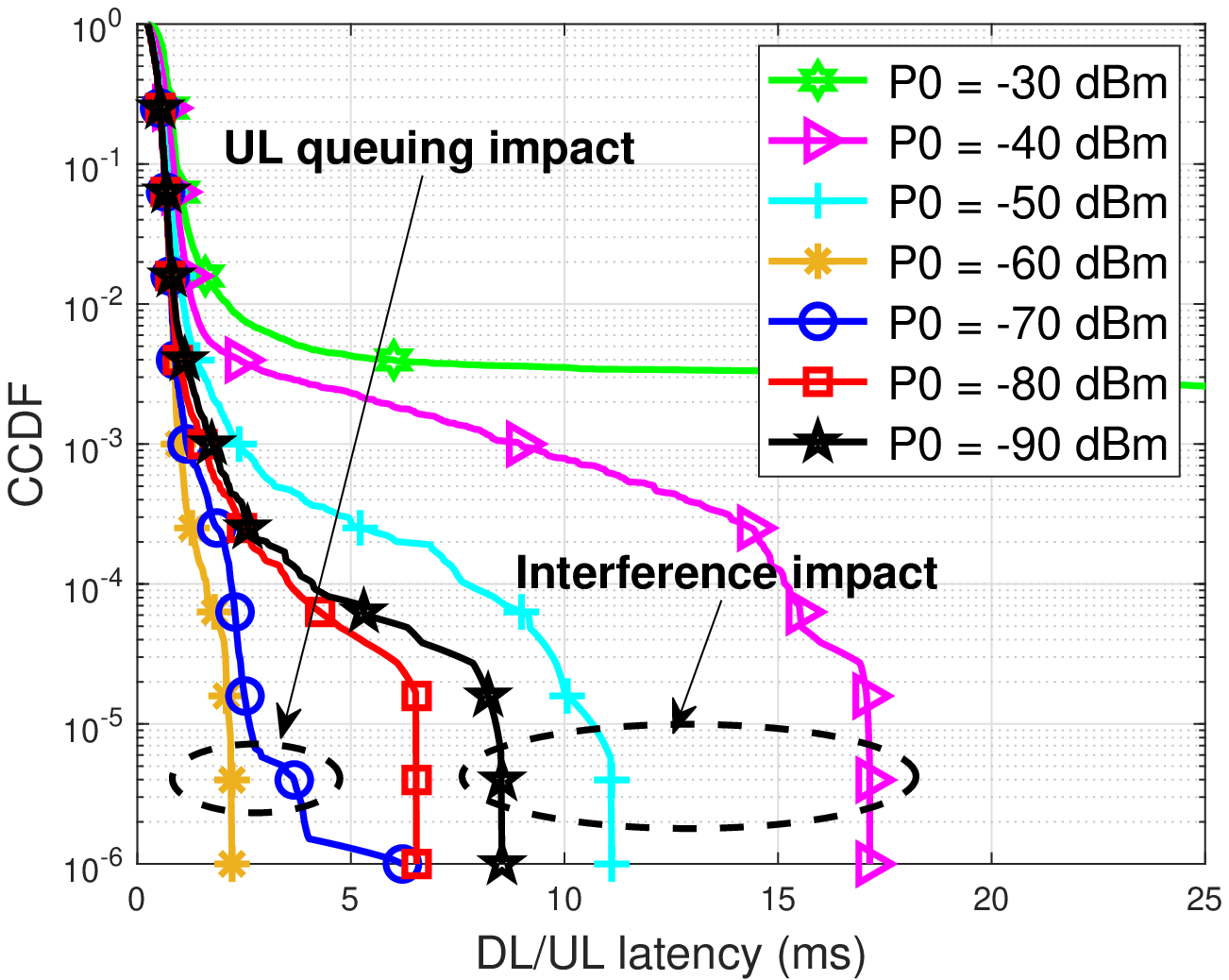}
\par\end{centering}
\centering{}{\small{}Fig. 2.}\textbf{\small{} }{\small{}Achievable
URLLC latency with dynamic TDD for different $P0$.}{\small \par}
\end{figure}

Fig. 3 depicts the CCDF of the achievable combined DL/UL URLLC latency
for different offered loads. For the low load region with $\varOmega=0.5$
Mbps, the URLLC target is achieved, i.e., a fully dynamic TDD satisfies
the URLLC outage latency target of 1 ms. This is mainly attributed
to the low CLI intensity and the smaller queuing delays under such
very low offered load. For higher offered loads of $\varOmega=3$
Mbps, the achievable URLLC outage latency inflicts a clear increase
due to the packet queuing delay. Moreover, the CLI is shown to have
a minor effect on the achievable URLLC latency for the low and moderate
offered load levels. 

\begin{figure}
\begin{centering}
\includegraphics[scale=0.6]{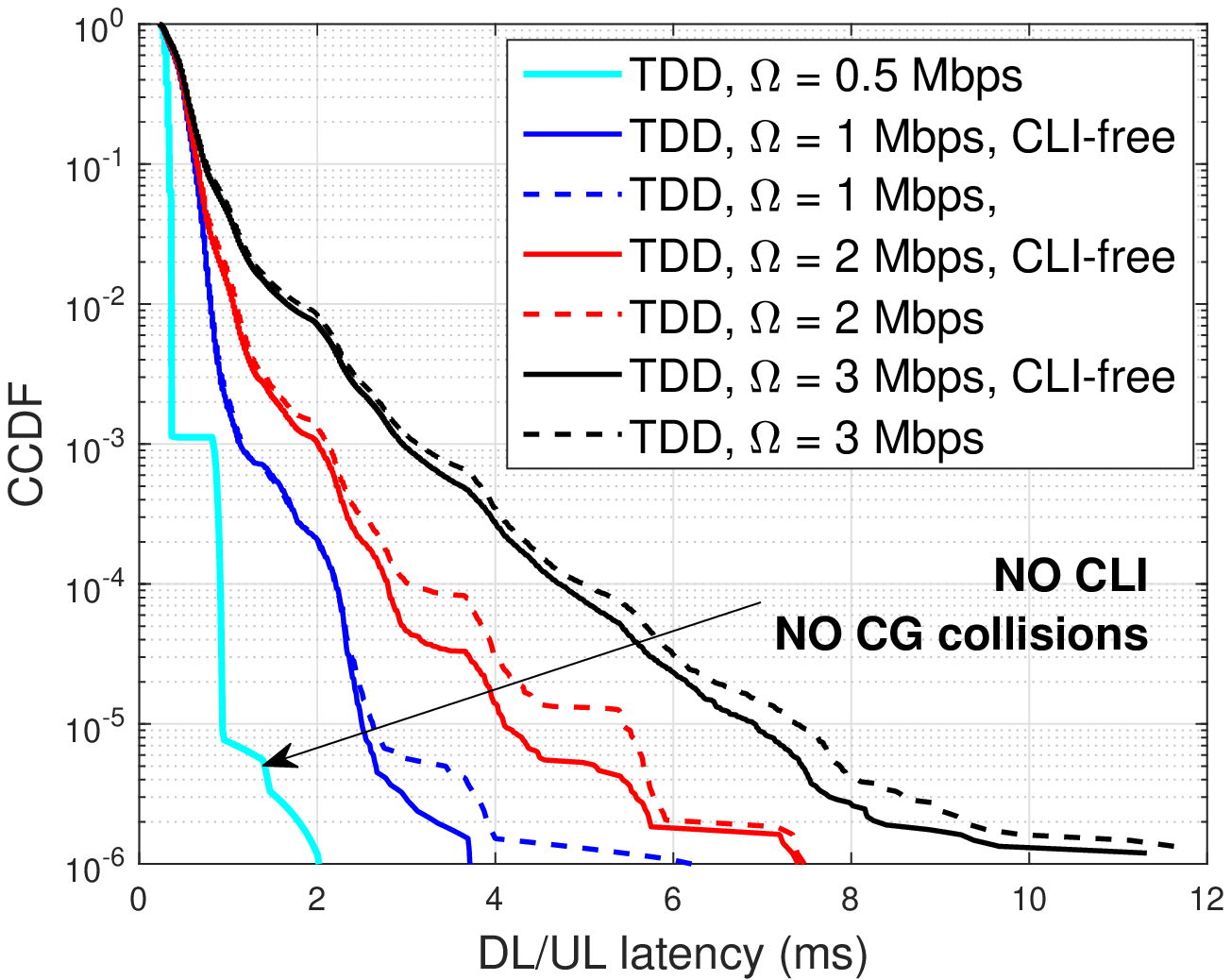}
\par\end{centering}
\centering{}{\small{}Fig. 3.}\textbf{\small{} }{\small{}Achievable
URLLC latency with dynamic TDD for different loads.}{\small \par}
\end{figure}

Next, we investigate the URLLC and eMBB coexistence performance under
different scheduling policies. Fig. 4 depicts the CCDF of the achievable
URLLC latency, where $\varOmega=2$ Mbps, and the eMBB traffic is
only incorporated in the DL directions with 3 eMBB UEs per cell, each
has a CBR of 0.5 Mbps. The throughput-based PF dynamic UE scheduler
fails to achieve a decent URLLC outage compared to the HoLD-aware
scheduler {[}24{]}. This is mainly because the latter considers the
latency statistics of pending URLLC UEs in the scheduling criterion.
It seeks to schedule the URLLC UEs with the largest HoLD statistics
while reducing the probability of the packet segmentation. Furthermore,
adopting a QoS-aware TDD link selection criterion tends to significantly
improve the achievable URLLC performance, i.e., 68\% outage latency
reduction compared to the URLLC QoS-unaware TDD selection criterion.
This is attributed to the fact that with the QoS-aware TDD link selection,
the selection of the TDD frame configuration is dictated by the URLLC
offered traffic size, instead of the aggregate URLLC/eMBB traffic,
reducing the TDD link switching delay of the urgent URLLC packets.

\begin{figure}
\begin{centering}
\includegraphics[scale=0.6]{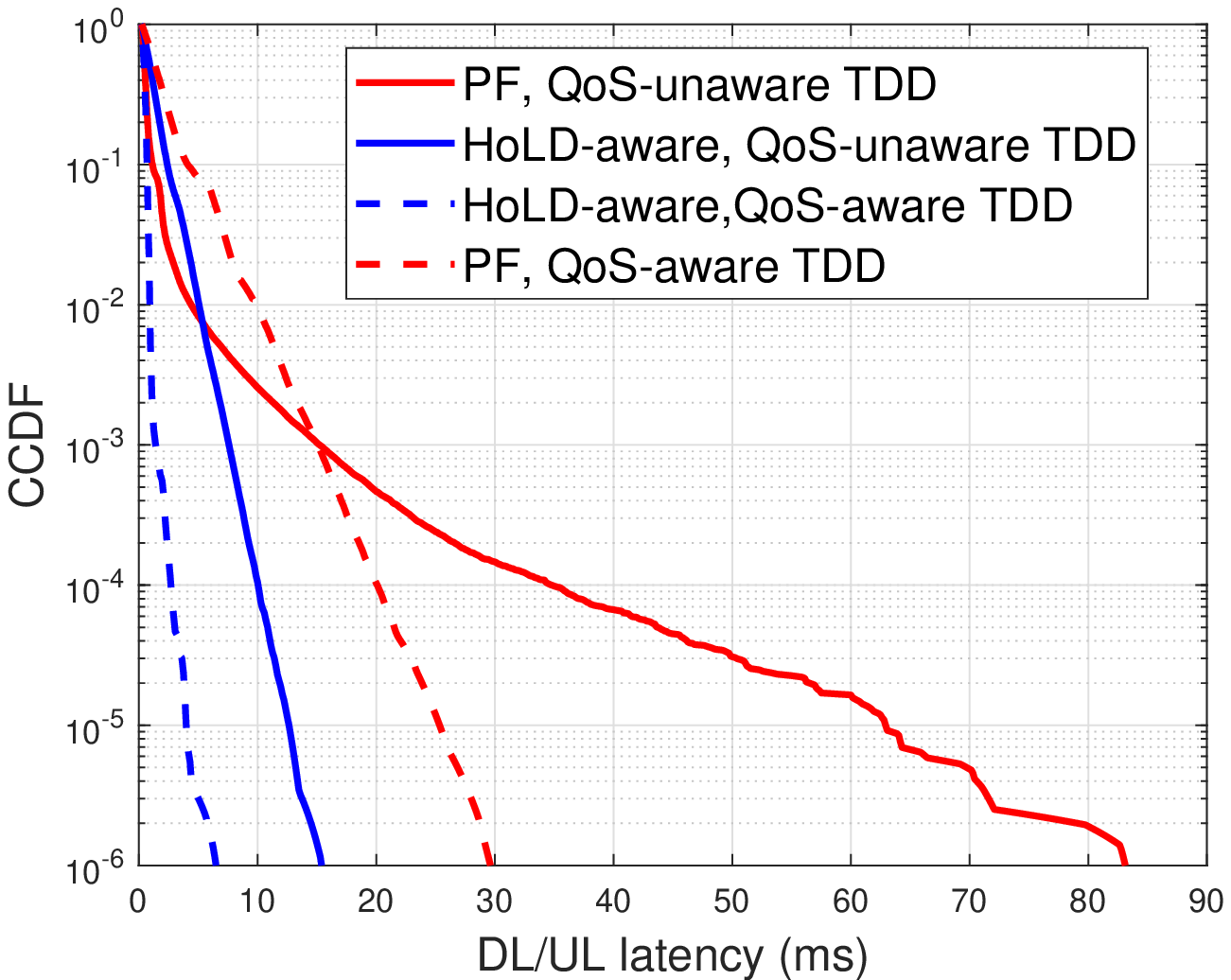}
\par\end{centering}
\centering{}{\small{}Fig. 4.}\textbf{\small{} }{\small{}Achievable
URLLC latency for different dynamic UE scheduling and TDD link selection.}{\small \par}
\end{figure}

Fig. 5 shows the empirical CDF (ECDF) of the achievable throughput
per eMBB UE. The source CBR rate is pre-configured as 0.5 Mbps per
UE. As depicted, the achievable eMBB throughput with the HoLD-aware
scheduler approaches the source CBR rate, while significantly outperforming
the case with the PF scheduler. This is mainly because: (1) the URLLC
transmissions are scheduled in a faster basis while the URLLC packet
segmentation is reduced, and (2) in the frequency domain, URLLC packets
are scheduled based on the throughput-to-average criterion which further
minimizes the required total number of PRBs to allocate the active
URLLC UEs. The HoLD-aware scheduler attempts to avoid the URLLC packet
segmentation, resulting from the insufficiently available free resources.
In case this is not possible, the scheduler seeks to inflict segmentation
of the URLLC packets that result in the lowest possible control overhead.
Therefore, it leaves more resources for the respective eMBB traffic,
and accordingly, achieving a highly optimized eMBB capacity compared
to the case with the PF scheduler. 

\begin{figure}
\begin{centering}
\includegraphics[scale=0.6]{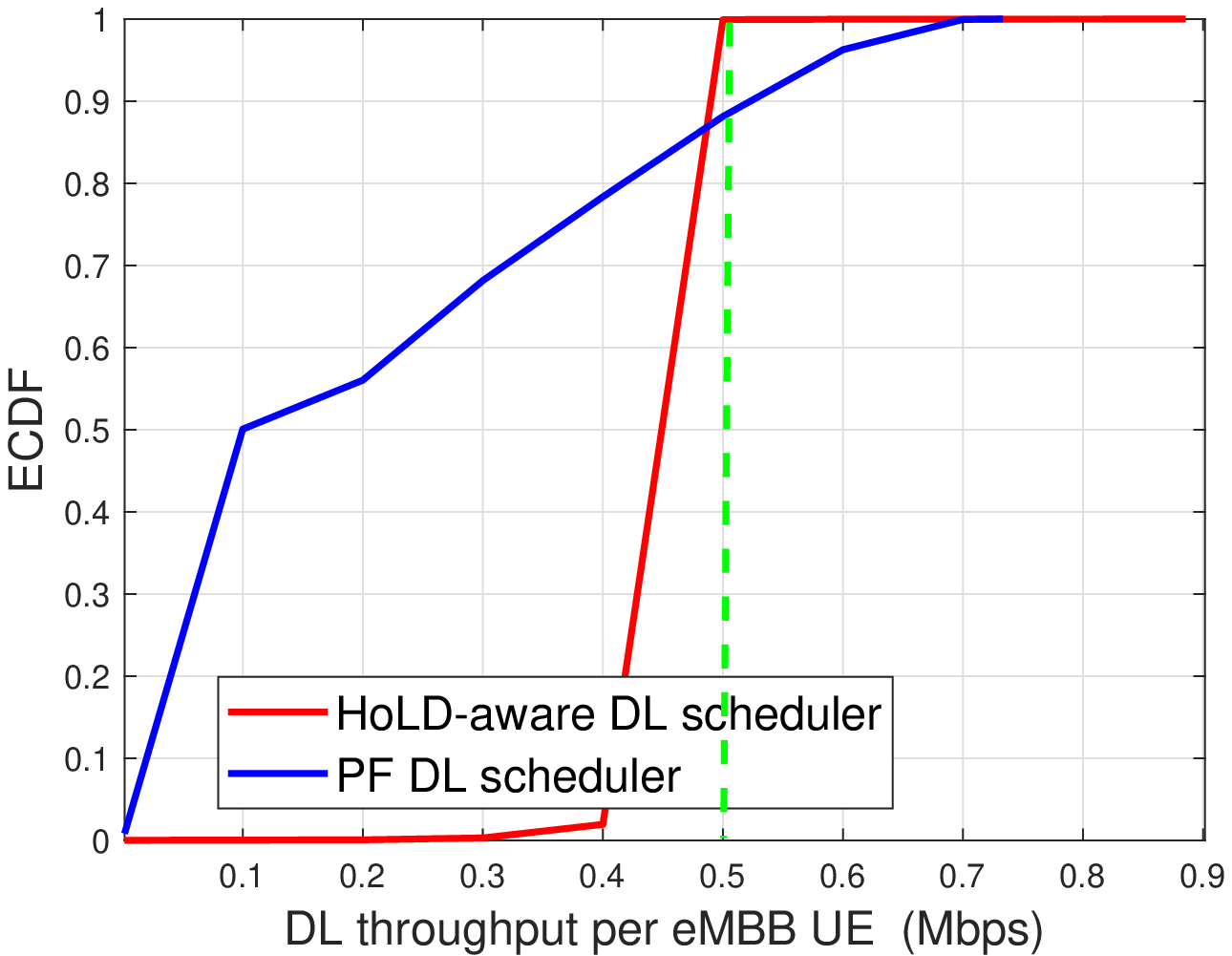}
\par\end{centering}
\centering{}{\small{}Fig. 5.}\textbf{\small{} }{\small{}Achievable
eMBB CBR rate for URLLC-eMBB coexistence.}{\small \par}
\end{figure}

Finally, we investigate the potential of the RL-based TDD frame selection
solution {[}19{]} compared to the non-RL based TDD frame selection
schemes, i.e., reactive TDD, where only the URLLC traffic is considered.
Fig. 6 depicts the achievable URLLC latency performance when such
learning approach is adopted for different load regions. As can be
observed, at the low load region, both the RL-based TDD and reactive
dynamic TDD schemes offer a similar URLLC outage performance. This
is mainly due to the low resource utilization, thus, adopting predefined
random UL/DL allocations during the selected TDD frame, in line with
the buffered traffic ratio, is sufficient. At the high load region,
the resource utilization increases, introducing additional queuing
delays for urgent URLLC packets in both the DL and UL directions.
Therefore, due to the smarter and latency-aware adaptation of the
RL-based TDD solution, the TDD learning approach obviously outperforms
the basic dynamic TDD scheme. 

\begin{figure}
\begin{centering}
\includegraphics[scale=0.7]{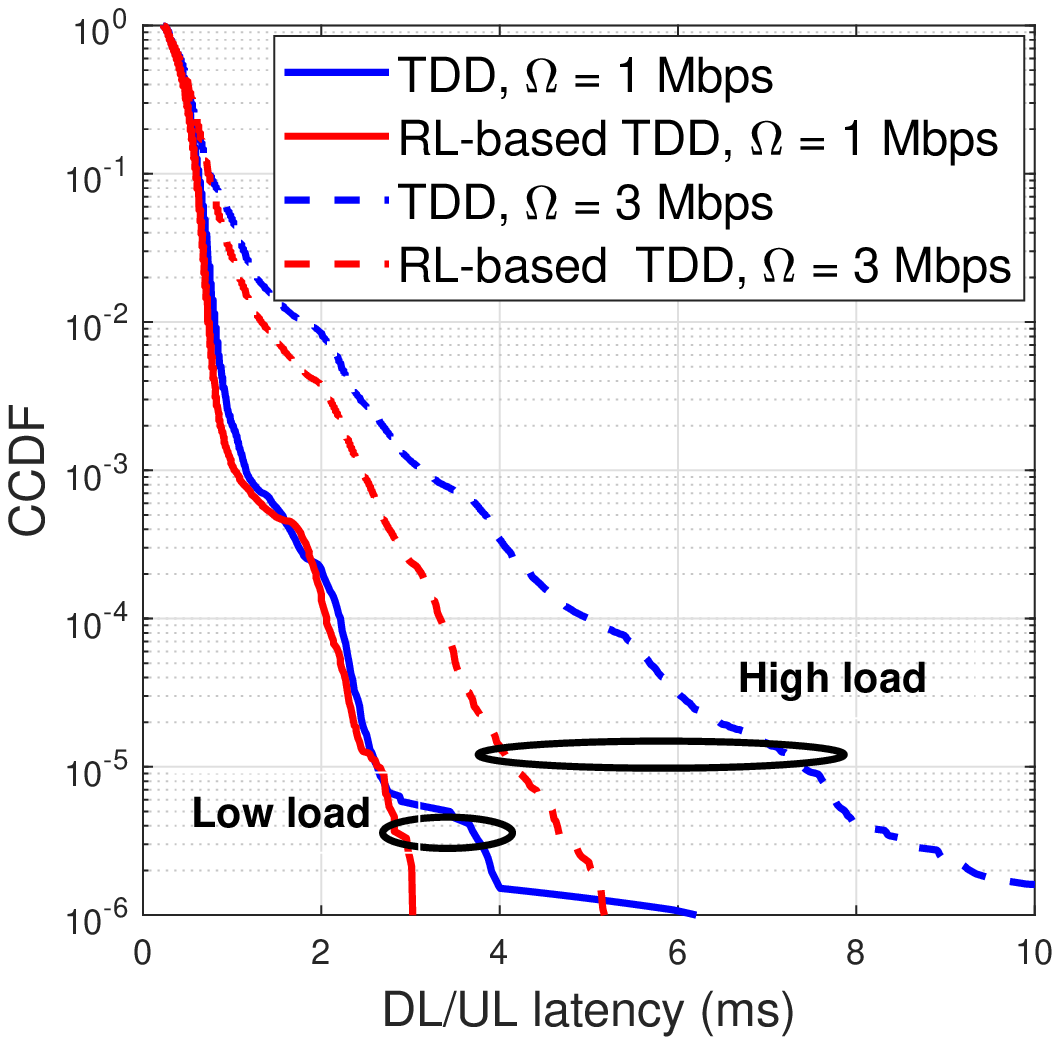}
\par\end{centering}
\centering{}{\small{}Fig. 6.}\textbf{\small{} }{\small{}Achievable
URLLC latency with the reinforcement learning approach.}{\small \par}
\end{figure}

\section{Concluding Remarks }

We have evaluated the achievable URLLC performance for the emerging
indoor factory automation 5G network deployments. We have analyzed
the state-of-the-art dynamic TDD duplexing scheme with optimized uplink
power control settings. For the URLLC-eMBB coexistence scenarios,
we adopt quality of service (QoS)-aware dynamic user scheduling and
TDD link selection strategy, respectively. The main recommendations
offered by this paper are as follows: (a) for the indoor factory deployments,
the CLI is not a major critical performance bottleneck as the case
with the macro networks, due to the InF prorogation conditions, and
the smaller difference between the uplink and downlink transmission
power, (2) the optimization of the uplink control settings has a vital
impact on the achievable URLLC outage performance. Unoptimized uplink
power control configurations could either lead to a further uplink
queuing delay or a significantly higher inter-cell same and cross-link
interference. Therefore, we recommend setting $P0=-61$ dBm within
the indoor factory deployments to achieve the best possible URLLC
outage latency, (3) within multi-QoS coexistence scenarios, latency-aware
dynamic user scheduling and TDD frame selection strategies are vital
to achieve a decent URLLC latency performance, and (4) reinforcement
learning (RL) based TDD frame adaptation is effective in achieving
a decent URLLC outage latency within InF deployments, through the
dynamic selection of the number and placement of the downlink and
uplink transmission opportunities across the TDD radio frame which
best reduces the overall radio latency. However, it requires a careful
modeling of the learning objectives, inputs, and outputs. 

\section{Acknowledgments}

This work is partly funded by the Innovation Fund Denmark \textendash{}
File: 7038-00009B. The authors would like to acknowledge the contributions
of their colleagues in the project.

\end{document}